# Loop Stirling Engines

**Minghao Deng**[a]

[a] *School of Energy Science and Engineering, Harbin Institute of Technology, Harbin, China*

**abstract**

The Stirling engine is a type of heat engine known as its high efficiency. It is applied in solar thermal power, cogeneration, space nuclear power, and other fields. Although there are many different types of Stirling engines, their airflow paths are always linear. This article designs two types of Stirling engines with loop airflow path: the O-type engines without regenerator and the 8-type engines with regenerator. The modeling and simulation of the O-type engines show its extremely excellent performance compared with the conventional Stirling engine. Because the regenerator is the main loss and power limitation in Stirling engines, O-type engines do not have this limitation. At the same time, its design without regenerator makes it more practical and has greater potential in terms of power. The 8-type engines use its unique 8-type airflow path to allow gas to enter the regenerator in advance, eliminating the almost useless four heat exchanges, resulting in higher thermal efficiency and better robustness.

| Nomenclature | |
|---|---|
| $c_p$ | specific heat at constant pressure (J/kg/K) |
| $c_V$ | specific heat at constant volume (J/kg/K) |
| $D$ | diameter (m) |
| $h$ | convective heat transfer coefficient (W/m²/K) |
| $H_1$ | enthalpy flowing out of the compression space (J) |
| $H_2$ | enthalpy flowing out of the expansion space (J) |
| $H_C$ | enthalpy flowing out of the cooler (J) |
| $H_H$ | enthalpy flowing out of the heater (J) |
| $H_R$ | enthalpy that cannot be fully regenerated (J) |
| $H_{high}$ | enthalpy of high-temperature airflow flowing into the regenerator (J) |
| $H_{low}$ | enthalpy of low-temperature airflow flowing into the regenerator (J) |
| $k$ | incomplete heat recovery rate defined by equation 18 (1) |
| $k_1$ | parameters defined by equation 3 (1/m) |
| $k_2$ | parameters defined by equation 4 (Pa²/K) |
| $\dot{m}$ | mass flow (kg/s) |
| $m_1$ | compression space gas quality (kg) |
| $m_2$ | expansion space gas quality (kg) |
| $p_0$ | inlet gas pressure (Pa) |
| $p_{aver}$ | average pressure (Pa) |
| $p_1$ | compression space gas pressure (Pa) |
| $p_2$ | expansion space gas pressure (Pa) |
| $R_g$ | gas constant (J/kg/K) |
| $S$ | heat exchange area (m²) |
| $T_1$ | compression space gas temperature (K) |
| $T_2$ | expansion space gas temperature (K) |
| $T_0$ | inlet gas temperature (K) |
| $T_C$ | temperature flowing out of the cooler (K) |
| $T_H$ | temperature flowing out of the heater (K) |
| $T_{in}$ | temperature of gas flowing into two spaces (K) |
| $T_W$ | wall temperature (K) |
| $T_R$ | temperature difference that cannot be fully regenerated (K) |
| $T_{high}$ | temperature of high-temperature airflow flowing into the regenerator (K) |
| $T_{low}$ | temperature of low-temperature airflow flowing into the regenerator (K) |
| $V_1$ | compression space gas volume (m$^3$) |
| $V_2$ | expansion space gas volume (m$^3$) |
| $\lambda$ | frictional drag factor (1) |
| $\gamma$ | heat capacity ratio (1) |
| $\eta_S$ | efficiency of conventional Stirling engine (1) |
| $\eta_8$ | efficiency of 8-type engine (1) |
| $\eta_R$ | efficiency of complete heat recovery (1) |

## 1. Introduction

The production and utilization of energy is a very important part of human production and life. Thermal conversion with high efficiency is an important topic in the field of energy. In recent years, Stirling engines have been applied as an efficient thermal conversion device in fields such as solar thermal power[1-4], cogeneration[5-7], and space nuclear power[8, 9], and other fields. Its reverse cycle can also be used as a refrigerator or heat pump[10].

The invention of the Stirling engine was based on the Stirling cycle. Although the thermodynamic processes in actual Stirling engines are much more complex than the ideal Stirling cycle[11-17], the ideal Stirling cycle still has reference value for studying Stirling engines. The ideal Stirling cycle is a closed gas thermodynamic cycle consisting of two adiabatic processes and two constant volume processes. Its thermal efficiency can reach the Carnot limit in theory. However, in reality, the Stirling engine's thermal efficiency is around 50% of the Carnot limit due to various losses. The regenerator is usually the mesh or porous structure, so the flow resistance of the regenerator is often high during high-speed engine operation. It limits the power of the Stirling engine.

There are three main classification methods for Stirling engines, including single or double acting, piston cylinder systems, and drive systems. Single acting means that only one piston is doing work, while double acting means at least two pistons are doing work. There are three main ways to arrange the piston cylinder system. Alpha type consists of two pistons in two cylinders respectively. Beta type refers to two pistons sharing a cylinder, and the back space of one piston is the front space of the other piston. Gamma type consists of two pistons each equipped with a cylinder, and the back space of one piston is connected to the front space of the other piston. The most diverse classification of Stirling engines is the drive system, which is mainly reflected in the material of the piston or the connection form between the piston and the output, including kinetic, free-piston[17], liquid piston, and thermoacoustic types[19, 20]. The kinetic type is the simplest drive system. It directly connects the piston to the output shaft through some geometric structures, such as crank-slider drive, rhombic drive[21, 22], swash-plate drive, etc. The kinetic type pistons have low degrees of freedom, while the free-piston type uses non rigid connection mechanisms to connect the pistons, giving them higher degrees of freedom and enabling them to work in more complex working conditions. The free-piston type has the advantages of self-starting, maintenance free, and long operating life. But the movement of the piston is more difficult to predict. It brings many difficulties to their design and analysis. The liquid piston type uses the liquid column as the piston. It actually belongs to the free-piston type. The thermoacoustic type uses the gas itself as the piston and also belongs to the free-piston type. But due to the significant differences in physical properties between gas and solid or liquid column, gas piston has some unique properties that make the thermoacoustic type stands out.

In summary, many different types of Stirling engines have been invented over the years. However, the path of airflow in the Stirling engine is always linear: gas flows back and forth between the compression space, cooler, regenerator, heater, and expansion space. This study aims to change the structure of the Stirling engine and add check valves to control the airflow path, allowing the airflow to flow in the loop-type. And do some research on them.

## 2. Design
### 2.1 Temperature analysis of conventional Stirling engines

The Stirling engine flows back and forth between the five components shown in Fig. 1 and converts thermal energy into kinetic energy. Designing a loop-type airflow path requires designing its temperature changes. The temperature change of high-efficiency thermal cycle needs to meet three conditions: absorbing heat from the heat source at high temperatures; Release heat to the cold source at low temperatures; The conversion between high and low temperatures should be as reversible as possible. Fig. 2 shows the temperature changes of Stirling engines, which can be caused by six processes: expansion or compression; Heating or cooling in the regenerator; Heating in the heater; Cooling in the cooler.

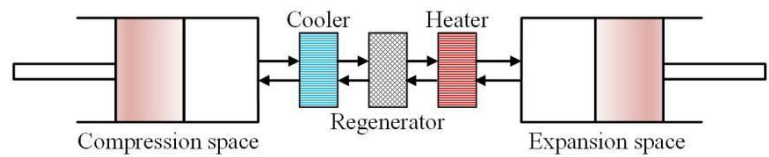

Fig. 1. The airflow path in conventional Stirling engines.

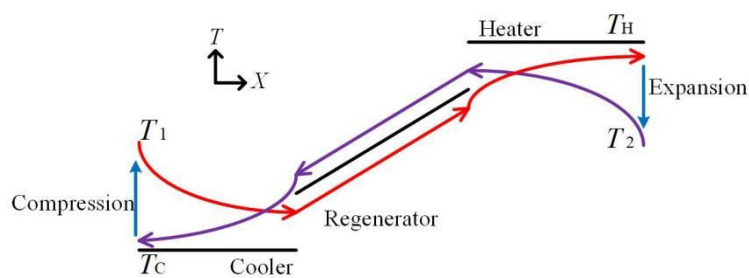

Fig. 2. Temperature changes in conventional Stirling engine cycles.

### 2.2 Design and analysis of the O-type engines
#### 2.2.1 Design of the O-type airflow path

It is difficult to directly construct a circular airflow path with a regenerator. Even without regenerator, there are four processes that can change the temperature of the gas. Attempt to construct a circular airflow path without regenerator as shown in Fig. 3, using check valves to control the airflow path. Fig. 4 shows its temperature variation. It can also perform a refrigeration cycle, as shown in Fig. 5.

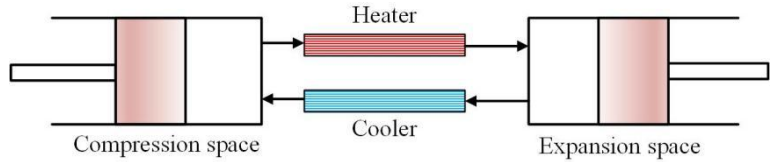

Fig. 3. The airflow path in O-type engines.

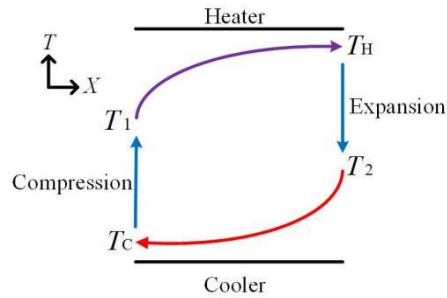

Fig. 4. Temperature changes in O-type engine cycles.

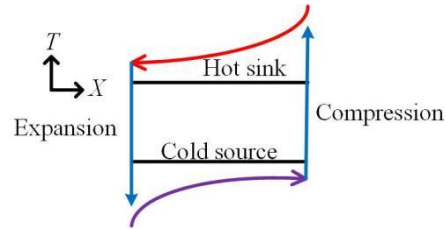

Fig. 5. Temperature changes in O-type refrigeration cycles.

In theory, the Brayton cycle can operate in this structure: the gas undergoes adiabatic processes in two spaces and isobaric processes in the heater and cooler. The actual thermodynamic process is more complex, and the performance of the engine needs to be simulated in order to obtain it. The regenerator is the main power limitation of conventional Stirling engines. However, simply abandoning the regenerator will directly connect the heater and cooler, resulting in a huge waste of energy and making it impossible to achieve high thermal efficiency. The design of the airflow path in the O-type engines cleverly solve this problem.

### 2.2.2 Classification of processes of the O-type engines

The thermodynamic process of the cycle is divided into the flow of gas in the heat exchanger (heater, cooler) and the inflow and outflow of two spaces (compression space, expansion space), as shown in Fig. 6.

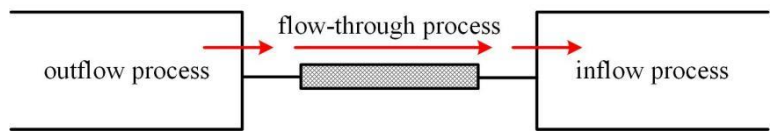

Fig. 6. Classification of thermodynamic processes.

The model is as follows: the gas is ideal gas; Two space insulation; The volume change function of the two spaces is determined and known; Neglecting the kinetic energy of gases; The change of airflow state in the heat exchanger is much faster than the change of gas state in the two spaces, which is enough to make the state of the airflow approach the steady flow in the current state of the two spaces. The steady flow assumption is completely accurate when the volume of the two spaces is much larger than that of the heat exchanger, but as the volume ratio of the two spaces to the heat exchanger decreases, the accuracy of this assumption also decreases.

From the model classification of Stirling engine [14, 15], this model has the characteristics of different level models. The steady flow assumption is an important assumption in this model: under the premise of tolerating some errors, the flow resistance and incomplete heat transfer can be calculated to evaluate the engine power. In fact, if add hysteresis correction and mass flow correction caused by pressure changes to the flow-through process model, most of the errors in the steady flow assumption can be removed.

### 2.2.3 The model of flow-through process

Assuming the heater and cooler are circular tubes. For the gas inside the tube at a certain moment, use steady-state flow analysis and ignore the kinetic energy term in the energy equation. Gas temperature and pressure vary along the tube:

$$\frac{dT}{T_W - T} = k_1 dx \tag{1}$$

$$dp + k_2 d\frac{T}{p} + \frac{\lambda k_2}{2D}\frac{T}{p} = 0 \tag{2}$$

Parameters $k_1$ and $k_2$:

$$k_1 = \frac{\pi h D}{\dot{m} c_p} \tag{3}$$

$$k_2 = \frac{16 R_g \dot{m}^2}{\pi^2 D^4} \tag{4}$$

Integrate to obtain the distribution of temperature and pressure along the tube. The pressure term in the denominator of the third term of equation (2) is approximated by the average pressure of the two spaces at this moment:

$$T = T_W + (T_0 - T_W)\exp(-k_1 x) \tag{5}$$

$$p - p_0 + k_2(\frac{T}{p} - \frac{T_0}{p_0}) + \frac{\lambda k_2}{2D p_{aver}}[T_W x + \frac{1 - \exp(-k_1 x)}{k_1}(T_0 - T_W)] = 0 \tag{6}$$

### 2.2.4 The model of inflow and outflow process

Analyze the process of gas flowing into two spaces, energy equation:

$$c_V d(mT) = c_p T_{in} dm - p dV \tag{7}$$

Combining equations such as the ideal gas state equation, solution:

$$dp = \frac{\gamma}{V}(R_g T_{in} dm - p dV) \tag{8}$$

$$dT = T(\frac{dp}{p} + \frac{dV}{V} - \frac{dm}{m}) \tag{9}$$

The outflow process is a simplified version of the inflow process. Just rewrite $T_{in}$ as the temperature of the gas inside the space. It is easy to infer that the gas inside the space undergoes adiabatic process during the outflow process. This can be mutually verified with equation (8).

### 2.2.5 The simulation of O-type engines

Write a python program to simulate O-type engine. When the current state of the two spaces is known, the next state of the two spaces can be obtained by calculating the inflow, outflow, and flow processes at each moment, and then the state changes of the two spaces in the entire cycle are simulated. Fig. 7 shows the basic steps of simulation. Fig. 8 shows the complete flowchart.

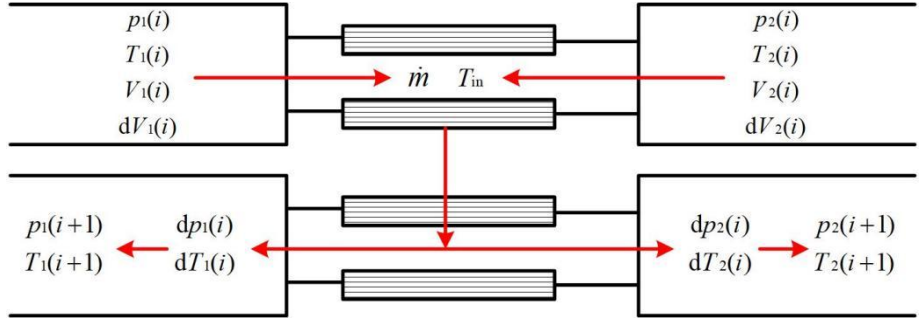

Fig. 7. The physical steps of simulation.

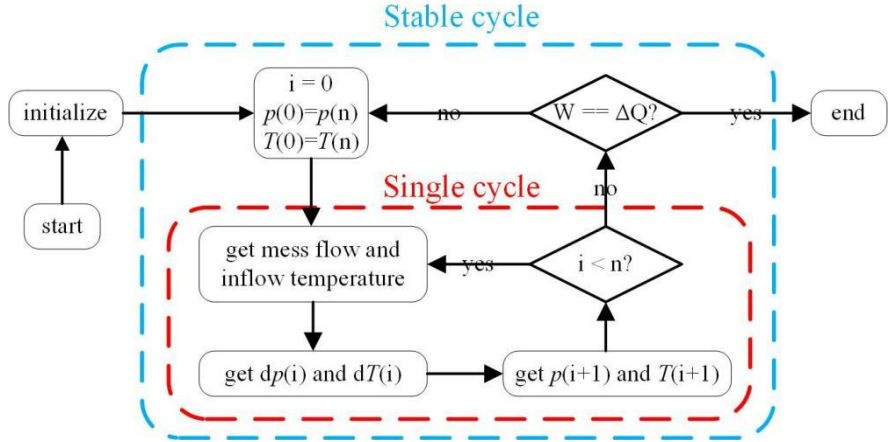

Fig. 8. Complete flowchart in the program.

Using the energy equations of two spaces as the criterion to determine whether the cycle is stable:

$$\int p_1 \mathrm{d}V_1 = \int_{\mathrm{d}m_1>0} c_p T_C \mathrm{d}m_1 - \int_{\mathrm{d}m_2>0} c_p T_1 \mathrm{d}m_2 \tag{10}$$

$$\int p_2 \mathrm{d}V_2 = \int_{\mathrm{d}m_2>0} c_p T_H \mathrm{d}m_2 - \int_{\mathrm{d}m_1>0} c_p T_2 \mathrm{d}m_1 \tag{11}$$

### 2.3 Design and analysis of the 8-type engines
### 2.3.1 Design of the 8-type airflow path

From the perspective of high and low temperature conversion, in O-type engines, adiabatic processes replace regenerator to complete the conversion between high and low temperatures. When considering large temperature differences, the temperature changes generated by the adiabatic process are not enough. Attempt to reactivate the regenerator to assist the gas in converting between high and low temperatures. The gas is regenerated after the adiabatic process and before the isobaric process to obtain the 8-type engine as shown in Fig. 9. The high and low temperature conversion is jointly completed by the adiabatic process and the regenerator, as shown in Fig. 10. Fig. 11 shows the temperature changes during the operation of the refrigeration cycle in the 8-type engines.

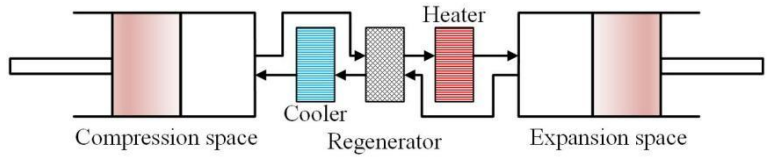

Fig. 9. The airflow path in 8-type engines.

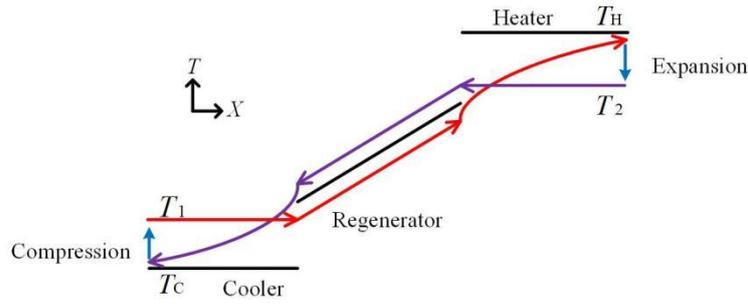

Fig. 10. Temperature changes in 8-type engine cycles.

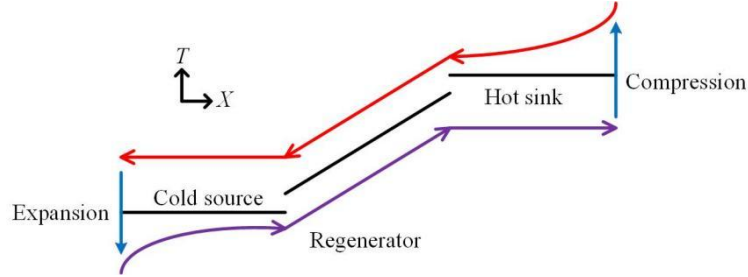

Fig. 11. Temperature changes in 8-type refrigeration cycles.

Compared to Fig. 2, the Fig. 10 eliminates four heat exchanges: after flowing out of the expansion space, heat is absorbed from the heater and immediately released to the regenerator; After flowing out of the compression space, heat is released to the cooler and immediately reabsorbed from the regenerator.

### 2.3.2 The preliminary analysis of 8-type engines

From both the airflow path and temperature changes, the 8-type engine is very similar to the Stirling engine. When studying the differences between the two, friction and incomplete heat transfer are mainly considered. The friction and incomplete heat transfer in the regenerator are much higher than those in the heater and cooler, which are the main factors limiting the performance of Stirling engines. Now suppose a Stirling engine is directly converted into an 8-type engine: changing the connection method of each part and adding check valves; The length of the heater and cooler has doubled. After this modification, the total friction of the airflow remained almost unchanged. The changes in incomplete heat exchange between the heater and cooler can also be ignored compared to those in the regenerator. The main research focuses on the impact of modification on incomplete heat transfer in the regenerator. So the model was simplified as: ignoring friction; Neglecting incomplete heat transfer of gas in heaters and coolers.

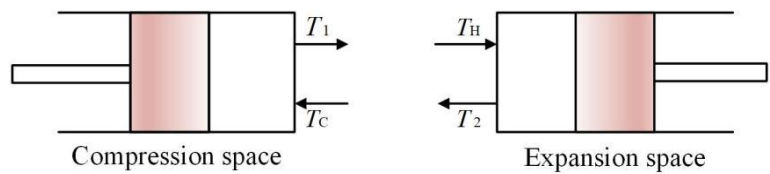

Fig. 12. The inflow and outflow model of two spaces.

The temperature of the gas flowing into and out of the two spaces before and after the modification is consistent, as shown in Fig. 12. This means that the state changes of the two spaces before and after the modification are completely consistent, so the work done is also completely consistent:

$$W = H_H - H_1 + H_C - H_2 \qquad (12)$$

$$H_H = \int_{dm_2>0} c_p T_H dm_2 = \int_{dm_1>0} c_p T_H dm_1 \tag{13}$$

$$H_1 = \int_{dm_2>0} c_p T_1 dm_2 \tag{14}$$

$$H_C = \int_{dm_2>0} c_p T_C dm_2 = \int_{dm_1>0} c_p T_C dm_1 \tag{15}$$

$$H_2 = \int_{dm_1>0} c_p T_2 dm_1 \tag{16}$$

Where $H_H$ is the enthalpy flowing out of the heater, $H_C$ is the enthalpy flowing out of the cooler, $H_1$ is the enthalpy flowing out of the compression space, and $H_2$ is the enthalpy flowing out of the expansion space.

Establish a heat transfer model for the regenerator as shown in Fig. 13. Among them, $T_R$ is the temperature difference of incomplete heat recovery, $T_{high}$ is the temperature of high-temperature incoming gas, and $T_{low}$ is the temperature of low-temperature incoming gas.

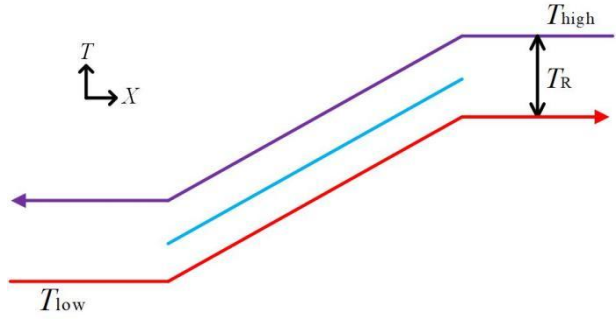

Fig. 13. Temperature distribution of regenerator.

Calculate the heat transfer power of the regenerator from different perspectives:

$$P_R = \frac{1}{2} T_R S h = (T_{high} - T_{low} - T_R) c_p \dot{m} \tag{17}$$

Define incomplete heat recovery rate:

$$k = \frac{T_R}{T_{high} - T_{low}} = \frac{H_R}{H_{high} - H_{low}} = \frac{1}{1 - \dfrac{Sh}{2\dot{m}c_p}} \tag{18}$$

Among them, $H_R$ is the enthalpy of incomplete heat recovery, $H_{high}$ is the enthalpy of high-temperature incoming gas, and $H_{low}$ is the enthalpy of low-temperature incoming gas. The four parameters of $S$, $h$, $\dot{m}$, and $c_p$ are the same before and after modification. Although the actual airflow situation is more complex, $k$ is still determined by these four parameters. It can be considered that the $k$ before and after the modification is the same.

The efficiency before modification is:

$$\eta_S = \frac{H_H - H_1 + H_C - H_2}{H_H - H_2 + k(H_H - H_C)} \tag{19}$$

The efficiency after modification is:

$$\eta_8 = \frac{H_H - H_1 + H_C - H_2}{H_H - H_2 + k(H_2 - H_1)} \tag{20}$$

The term with $k$ refers to the increase in heat absorption caused by incomplete heat recovery. The efficiency when fully reheated is:

$$\eta_R = \frac{H_H - H_1 + H_C - H_2}{H_H - H_2} \tag{21}$$

Comparison reveals:

$$\frac{1}{\eta_S} - \frac{1}{\eta_8} = k\left(\frac{2}{\eta_R} - 1\right) \tag{22}$$

This indicates that from the perspective of incomplete heat transfer in the regenerator, the 8-type airflow path can reduce the additional heat absorption caused by incomplete heat recovery by reducing the temperature at both ends of the regenerator, thereby improving efficiency. The less sufficient the heat recovery (the larger the $k$) or the lower the thermal efficiency at ideal complete heat recovery (the smaller the $\eta_R$), the more significant the efficiency improvement. This means that the robustness of the 8-type engine is better.

## 3. Results and Discussion
### 3.1 The O-type engine designed under GPU-3 engine conditions

Taking General Motors' GPU-3 engine [22-26] as the comparative object, the performance calculation, parameter design, and optimization of the O-type engine were carried out under the same operating conditions. Table 1 shows the main design requirements. The maximum total volume of gas in the O-type engine is set to 350cm$^3$, which is close to GPU-3. In order to make the model more accurate, the total volume of the two spaces is set to be 10 times the total volume of the heater and cooler.

Table 1

Design specifications

| Heater temperature | 977K | Working gas | Helium |
|---|---|---|---|
| Cooler temperature | 288K | thermal conductivity | 0.3445W/m/K(977K) |
| Average pressure | 2.76MPa |  | 0.1501W/m/K(288K) |
| Heater diameter | 3.02mm | dynamic viscosity | 44.13*10$^{-6}$kg/m/s(977K) |
| Cooler diameter | 1.09mm |  | 19.23*10$^{-6}$kg/m/s(288K) |

In order to verify the reliability of the program, the GPU-3 engine was also be simulated. Table 2 shows the comparison between simulation results and experimental values. The regenerator in the simulation program is set to fully recover heat without friction, so there is a significant difference from the experimental results[24, 25]. But compared to the adiabatic model[23, 24, 26], it takes into account the losses of the heater and cooler, so its performance is between the adiabatic model and experimental values. And because the losses of the regenerator are greater than those of the heater and cooler, simulation's performance is closer to the adiabatic model. The simulation of GPU-3 has to some extent demonstrated the reliability of the simulation program.

Table 2

Performance of GPU-3

|  | Adiabatic model | Simulation program | Experiment |
|---|---|---|---|
| Efficiency(%) | 62.3 | 49.7 | 21.3 |
| Power(kW) | 8.30 | 7.08 | 2.42 |

Designed three O-type engines with different power levels. Table 3 shows their design parameters and simulation results. In order to achieve higher thermal efficiency, the number of tubes in the heat exchanger should be reduced, the length of it should be increased, and the cycle frequency should be lowered. The performance of GPU-3 in simulation is slightly better than O-type engine. Because it has an additional ideal regenerator and can always use both heater and cooler at the same time.

Table 3

Optimized design parameters and performance of the engine

|  | O-1 | O-2 | O-3 | GPU-3 |
| --- | --- | --- | --- | --- |
| Number of heater tubes | 32.8 | 19.1 | 14.6 | 40 |
| Number of cooler tubes | 407 | 259 | 207 | 312 |
| Heater length(cm) | 8.14 | 14.5 | 19.4 | 24.5 |
| Cooler length(cm) | 3.34 | 4.97 | 5.94 | 4.61 |
| Cycle frequency(Hz) | 175 | 92.7 | 62.8 | 41.7 |
| Simulation efficiency(%) | 36.7 | 42.0 | 44.4 | 49.7 |
| Simulation power(kW) | 9.95 | 7.77 | 6.25 | 7.08 |

The simulation program has poor accuracy in simulating GPU-3. But it is a program developed for O-type engines. It can be considered that it has a relatively accurate simulation for O-type engines. So comparing the simulation results of the O-2 engine with the experimental values of the GPU-3 engine, the thermal efficiency is about twice and the power is about three times. The actual O-type engine has some losses that were not investigated in the program, such as gas leaks. But they are not the main losses of the engine and will not cause too much inaccuracy in the simulation results.

Draw temperature change diagrams and PV diagrams for two cycles for analysis, as shown in Fig. 14. It can be seen from (a) and (c) that the cycle logic of the two engines is different. The temperature of both engines varies periodically. In the GPU-3 engine, gas expands and cools in the expansion space and then enter the heater to absorb heat. But in the O-type engine, the gas directly enter the cooler for cooling after expanding and cooling in the expansion space. The O-type engine absorbs heat at high temperatures and releases heat at low temperatures, which is opposite to the GPU-3 engine. The highest temperature of the gas in the expansion chamber of the O-type engine is significantly lower. And from (b) and (d), it is found that the highest pressure and compression ratio of the O-type engine are significantly higher. Because it requires more adiabatic processes to complete the transition between high and low temperatures. In order to reduce the heat exchange temperature difference, it must have a higher compression ratio.

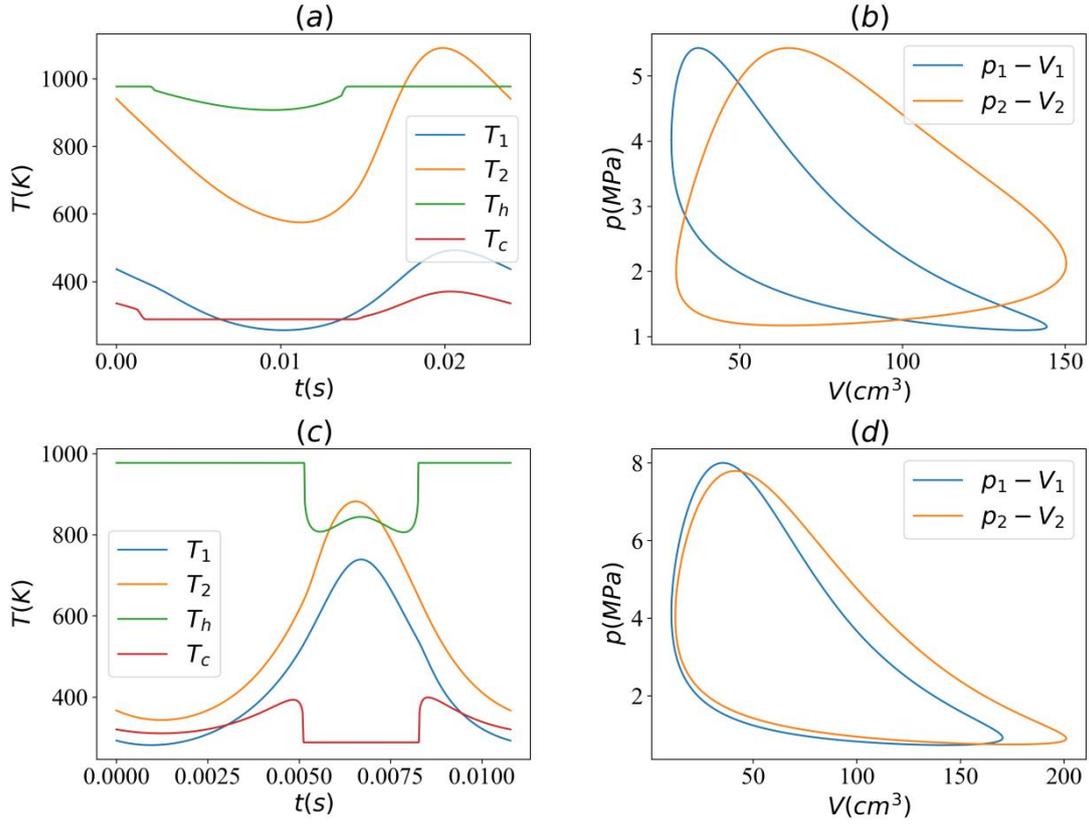

Fig. 14. Temperature change curves and PV diagrams of GPU-3 engine and O-2 engine.

## 3.2 Other operating condition tests

Performance calculations, parameter design, and optimization were conducted on the O-type engine under three different operating conditions. The three operating conditions are: Thermal conversion of high-performance engines to middle temperature heat sources; thermal conversion of low-cost engines to low temperature heat sources; Household refrigerators such as air conditioners and freezers. The working gas uses nitrogen gas. The total volume of the two spaces is set to 10 times that of the heater and cooler. The maximum total volume of gas is set to 350cm$^3$. Table 4 shows four set values and two simulation results for three operating conditions. When optimizing parameters, a function of power density and relative efficiency is used to evaluate the quality of the cycle: Power density is the ratio of power to the volume of the working fluid; Relative efficiency is the ratio of thermal efficiency to Carnot efficiency. After optimization, the thermal efficiency under all three operating conditions exceeded 50% of the Carnot limit.

Table 4

Setting values and simulation results for three operating conditions

|  | Heat source temperature | Cold source temperature | Maximum pressure | Heat exchange diameter | Power density | Relative efficiency |
|---|---|---|---|---|---|---|
| Middle temperature difference | 600K | 300K | 5MPa | 1mm | 4.48kW | 57.8% |
| Low temperature difference | 400K | 300K | 1MPa | 10mm | 5.48W | 55.3% |
| Household refrigeration | 300K | 270K | 1MPa | 10mm | 8.43W | 52.1% |

Fig. 15 shows the temperature change and PV diagrams of the gas during cyclic stability in the simulation program. Gases tend to absorb heat from the heat source at higher temperatures and release heat to the cold source at lower temperatures, as shown in (a), (c). This can reduce the heat exchange temperature difference, achieve lower entropy increase, and make the cycle more efficient. In the refrigeration cycle, while reducing the heat transfer temperature difference, it is also necessary to possibly minimize negative heat transfer as shown in (e). The gas in the compression space ($p_1$-$V_1$) runs counterclockwise, and the gas in the expansion space ($p_2$-$V_2$) runs clockwise, as shown in (b), (d), (f). The total work done by the gas is positive as shown in (b), (d). The total work done by the gas is negative as shown in (f).

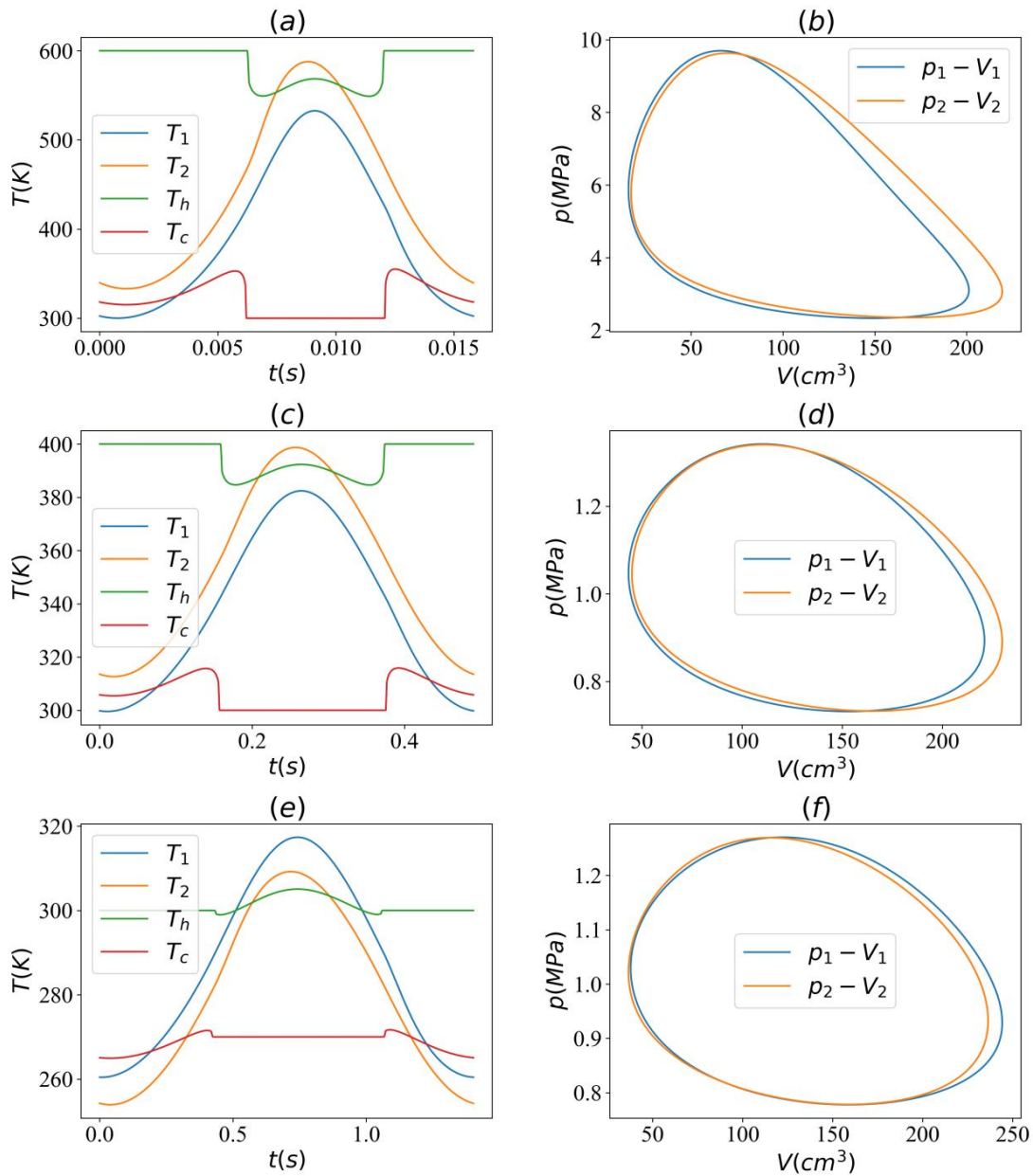

Fig. 15. Temperature change curves and PV diagrams under different operating conditions.

More precise control of the gas process can improve thermal efficiency, but it will reduce the utilization rate of the heat exchanger. The O-type and 8-type engines is exactly like this: it never uses both the heater and the cooler at the same time. Multiple thermal cycles sharing a heat exchanger and controlling various gas valves through a circuit system may be a better solution.

## 4. Conclusion

This article designs Stirling engines with O-type and 8-type airflow paths based on reasonable temperature changes. Modeling the O-type engines including flow-through process model, inflow and outflow process model. A simulation program for O-type engine was developed using Python language based on the model. Optimization of O-type engines under different operating conditions was carried out based on simulation programs. The results of the program simulation indicate that:

1. Compared to the GPU-3 Stirling engine under the same operating conditions, the O-type Stirling engine has twice thermal efficiency and three times power.

2. O-type engines with efficiency exceeding 50% Carnot limit can be designed under various temperature heat source thermal conversion and household refrigeration conditions.

From the process of modeling and simulation, it can also be seen that the structure without regenerator makes the theoretical research of O-type engines simpler and its manufacturing cost lower. All of these give it higher practicality than conventional Stirling engines. In addition to the O-type engine, this article also conducted some preliminary analysis on the 8-type engine comparing with the conventional Stirling engine. The 8-type engines have higher thermal efficiency and better robustness then conventional Stirling engines. The cost is increasing the length of the heater and cooler.

## Postscript-2025.04.06

Given the outstanding performance of the loop Stirling engine, all current Stirling engine researchers should shift their research to the loop Stirling engine. Except for thermoacoustic engines, almost all Stirling engines and refrigerators can be manufactured into loop structures with better performance, especially O-type structures. Time will prove my point.

Because the current academic trend of being flashy but not practical, I have given up submitting this paper to a journal, because I don't want to waste time on meaningless content filling and image beautification. So-called: 'bad money drives out good money'. I think I have done enough research on this topic and don't want to delve deeper. And I have recently become interested in some other topics, so I will not continue to research this topic. My final effort for this research is to publish it on arxiv and share it with other researchers in the field of Stirling engines.

During the process of sharing this paper, I discovered that Luo from Hunan University has recently independently designed the loop Stirling engines[27-30]. I researched the O-type engine before him in 2022. He conducted research on the 8-type engine before me.

## Postscript-2025.04.11

Many researchers have pointed out that check valve in loop structures cannot operate at high frequencies, so I have conducted some investigations on this question. I found that most car engines use the valve distributing mechanism to control the airflow path. So the control of the airflow path in the loop structure does not necessarily rely on check valves, and perhaps use the valve distributing mechanism. Furthermore, as gas thermal conversion devices using piston cylinder systems, Stirling engines may have many aspects to learn from more mature internal combustion engines.

Due to the limitations of the model and program, there is definitely a certain degree of inaccuracy in the simulation results. In the optimization process of loop engines, the inaccuracy of simulation results is definitely biased towards high-performance direction. Many losses in actual engines are also not included in the model, including the negative effects caused by controlling the airflow path. But this was not intentionally caused by me. In order to make the model more accurate, I even sacrificed some performance (the volume setting). But the model targets the main performance limitation: incomplete heat transfer and flow resistance. Unless controlling the airflow path is truly difficult to achieve, the results should not be completely unrealistic.

## 一些拙见

循环中熵增越少，循环的效率越高：

$$\eta = 1 - (\frac{\Delta S}{Q_H} - \frac{1}{T_H})T_C$$

所以为了获得高效率的循环，一定要尽量减少一切熵增。然而气体从热源吸热以及向冷源放热，这两个过程中的熵增是不可避免的。换热过程中的熵增与换热温差成正比，进而与换热功率成正比，进而与输出功率正相关。所以对于热力循环而言，功率和效率永远是矛盾的。在最完美的状态下，除了吸热和放热以外没有其他熵增，式（24）成立：

$$P = \frac{T_H + T_C - (1-\eta)T_H - \frac{1}{1-\eta}T_C}{R_H + R_C} \leq \frac{(\sqrt{T_H} - \sqrt{T_C})^2}{R_H + R_C}$$

所以将吸热和放热过程中的熵增视为有效熵增，将其他熵增视作无效熵增。当然，有效熵增也存在一个限度。观察上式，不等号成立的情况有特殊意义：此时功率达到最大值，对应的吸热以及放热温度应为临界温度，此时继续增加换热温差也不能获得更大的功率。临界温度取决于两端热阻，然而却总不会越过$\sqrt{T_H T_C}$，差也不会少于$\sqrt{T_H} - \sqrt{T_C}$。气体要在高于热临界温度时吸热，在低于冷临界温度时放热，但如果只是这样，两端的气体各行其事，无法构成一个循环。也就是说，还需要高低温之间的转换，熵增较少的方法是使用绝热过程或回热器。而考虑转换温差，在效率确定时，发现转换温差与吸热温度成正比。于是为了降低高低温转换的难度，通常会降低吸热温度，表现到发动机的设计上就是热端热阻应当高于冷端热阻。再考虑对外做功，气体的体积也要形成循环，所以气体的压缩和膨胀过程也是必不可少的。对于理想斯特林循环而言，压缩和膨胀过程与换热过程结合在一起表现为等温过程，高低温转换则由回热器承担。对于理想布雷顿循环而言，换热过程表现为等压过程，压缩和膨胀过程与高低温转换结合在一起表现为绝热过程。实际斯特林发动机中存在的问题在于，气体压缩和膨胀时产生的温度变化并没有被合理放置于高低温转换之中，而是重新流回了加热器和冷却器之中。当然这样做也有好处，那就是没有浪费加热器和冷却器的使用时长，相当于降低了热阻。所以下一步的优化可以恢复换热器的使用时长，比如将多个气流路径与多个换热器复合使用。加热器和冷却器个数之和等于气流路径个数的两倍，或者使单个气流路径包括多个加热器和冷却器，总之使每个加热器和冷却器不停用。

## 额外内容

众所周知，斯特林发动机的功率和效率一般是负相关的，而它们之间的关系并没有被量化表示。虽然仿真手段能得到功率-效率的相关曲线，但那并非一般性的关系。尝试使用较为初等的方程量化功率与效率之间的关系。图 16 展示了包括热阻 $R_H$ 和 $R_C$ 的简单模型。假定模型中除了图中两个换热以外全是可逆过程。

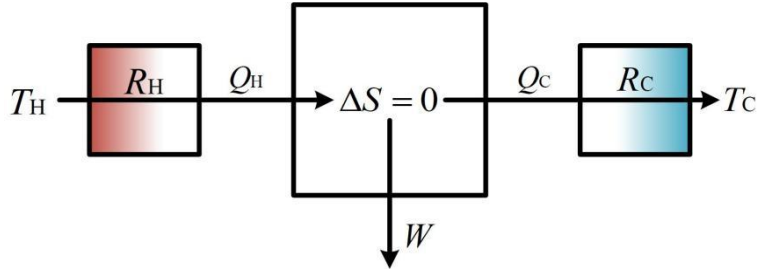

图 16

熵增为零时：

$$\Delta S = \frac{Q_C}{T_C + fR_C Q_C} - \frac{Q_H}{T_H - fR_H Q_H} = 0 \tag{23}$$

式中 $f$ 是循环频率，容易解出：

$$P = \frac{\eta T_H - \frac{\eta}{1-\eta} T_C}{R_H + R_C} = \frac{T_H + T_C - (1-\eta)T_H - \frac{1}{1-\eta}T_C}{R_H + R_C} \leq \frac{(\sqrt{T_H} - \sqrt{T_C})^2}{R_H + R_C} \tag{24}$$

特别的，对于 O 型发动机和 8 字型发动机而言，加热器和冷却器并不能同时工作。假设 $k_R$ 为加热器的工作时间占比，那么式（24）将被修正为：

$$P = \frac{\eta T_H - \frac{\eta}{1-\eta} T_C}{\frac{R_H}{k_R} + \frac{R_C}{1-k_R}} = \frac{T_H + T_C - (1-\eta)T_H - \frac{1}{1-\eta}T_C}{R_H + R_C + \frac{1-k_R}{k_R}R_H + \frac{k_R}{1-k_R}R_C} \leq \frac{(\sqrt{T_H} - \sqrt{T_C})^2}{(\sqrt{R_H} + \sqrt{R_C})^2} \tag{25}$$

热阻增大到了原来的两倍左右，这是因为换热器的利用率降低了。

这一模型虽然简单但是有许多局限性。首先，由于模型中并没有引入回热器的熵增，这一模型更适用于 O 型发动机。其次，由于没有气流与管壁的摩擦带来的熵增描述，随着发动机转速的增加，对流换热的热阻会被无限减小，直到只剩管壁的热阻。接下来尝试向模型中加入摩擦损失的表述。

最简单的表述直接由式（25）改写：

$$P = \frac{\eta T_H - \frac{\eta}{1-\eta} T_C}{\frac{R_H}{k_R} + \frac{R_C}{1-k_R}} - fF_C - fF_H \tag{26}$$

其中 $F_C$ 和 $F_H$ 是单次循环中冷却器和加热器中因摩擦损失的功。但这样没有考虑摩擦产生的热量对内部的影响，更精确的模型应当是计算摩擦带来的熵增并改写式（23）：

$$\Delta S = \frac{Q_C}{T_C + \frac{fR_C Q_C}{1-k_R}} - \frac{Q_H}{T_H - \frac{fR_H Q_H}{k_R}} = \frac{F_C}{T_C + \frac{fR_C Q_C}{1-k_R}} + \frac{F_H}{T_H - \frac{fR_H Q_H}{k_R}}$$
$$\frac{Q_C - F_C}{T_C + \frac{fR_C Q_C}{1-k_R}} - \frac{Q_H + F_H}{T_H - \frac{fR_H Q_H}{k_R}} = 0 \tag{27}$$

这个方程给出了 $Q_C$ 和 $Q_H$ 的关系，只要确定了换热器和气体工作时的相关参数，进而可以得到 $P$ 和 $\eta$ 的关系以及发动机这种工况下的功率极限。事实上它是一个关于 $P$ 的二次方程，但是它的解不简洁，所以不在这里给出。

定熵模型为 O 型斯特林发动机的研究提供了一种不依靠仿真的方法。它模糊了气体循环的实际变化过程，根据熵理论表述了气体在换热器中的不完全换热和流阻带来的熵增。与讨论具体的理想热力循环相比，只少了流入空间的气体与空间内气体混合时，因温度不一样造成的熵增。是一个可以粗略估计发动机性能的简单模型。考虑到模型本身就很粗糙，没有使用式（27）的必要，使用式（26）即可。